\documentclass[aps,prd,nofootinbib,twocolumn,superscriptaddress]{revtex4}

\usepackage{graphicx}
\usepackage{amsmath}
\usepackage{slashed}
\usepackage{color}

\newcommand{\ie}{{\it i.e. }}

\newcommand{\be}{\begin{equation}}
\newcommand{\ee}{\end{equation}}
\newcommand{\br}{\begin{eqnarray}}
\newcommand{\bea}{\begin{eqnarray}}
\newcommand{\eea}{\end{eqnarray}}
\newcommand{\er}{\end{eqnarray}}
\newcommand{\ba}{\begin{array}}
\newcommand{\ea}{\end{array}}
\newcommand{\bi}{\begin{itemize}}
\newcommand{\ei}{\end{itemize}}
\newcommand{\bn}{\begin{enumerate}}
\newcommand{\en}{\end{enumerate}}
\newcommand{\bc}{\begin{center}}
\newcommand{\ec}{\end{center}}

\newcommand{\beq}{\begin{equation}}
\newcommand{\eeq}{\end{equation}}

\newcommand{\gsim}{\lower.7ex\hbox{$\;\stackrel{\textstyle>}{\sim}\;$}}
\newcommand{\lsim}{\lower.7ex\hbox{$\;\stackrel{\textstyle<}{\sim}\;$}}

\def\mysection#1{{\bf #1.} }

\begin{document}

\title{Testing Right-Handed Currents at the LHC}

\author{Matti Heikinheimo}
\affiliation{National Institute of Chemical Physics and Biophysics, R\"avala 10, 10143 Tallinn, Estonia}

\author{Martti Raidal}
\affiliation{National Institute of Chemical Physics and Biophysics, R\"avala 10, 10143 Tallinn, Estonia}
\affiliation{Institute of Physics, University of Tartu, Estonia}

\author{Christian Spethmann}
\affiliation{National Institute of Chemical Physics and Biophysics, R\"avala 10, 10143 Tallinn, Estonia}

%\email[]{martti.raidal@cern.ch}

\date{\today}

\begin{abstract}
The CMS Collaboration has published two different searches for new physics that contain possible hints for excesses in $eejj$ and $e\nu jj$ final states.
Interpreting those hints as a possible signal of a right-handed gauge boson $W_R$ with mass 2-2.5~TeV may have profound implications 
for our understanding of the gauge structure of nature and Grand Unification, the scalar sector accessible at the LHC, neutrino physics, and the baryon asymmetry of the Universe.
We show that this interpretation is, indeed, consistent with all existing constraints. However, before making premature claims we propose a number of cross-checks  at the LHC14
that could confirm or falsify this scenario. Those include searches for a $Z_R$ resonance and the related new scalar sector around 6-7~TeV. Additionally, large effects in top-quark spin-asymmetries in single top production are possible. 
\end{abstract}

\maketitle

\section{Introduction}

In dedicated searches for right-handed currents~\cite{Khachatryan:2014dka} the CMS Collaboration at the LHC at CERN 
claims to see  a $2.8\sigma$ deviation from the Standard Model in the channel  
\[ pp \to 2j + e e ,\]
which is peaked around an invariant mass of the 4 objects of 2-2.5 TeV. 
The corresponding excess has not been observed in the $pp \to 2j + \mu \mu$ channel. 
At the same time, $2.4\sigma$ and $2.6\sigma$ excesses in the final states $2j+e e$ and $2j+e \nu$, respectively, 
have also been observed in searches for leptoquarks~\cite{Khachatryan:EXO12041} in which other kinematical
variables ($j\ell$ invariant masses) have been reconstructed. 
It is intriguing to argue that both excesses may have the same new physics origin.

In the pair production of leptoquarks~\cite{Khachatryan:EXO12041}, one expects to see no clear peak in the four particle invariant mass distribution, disfavouring leptoquarks as an explanation for the excess observed in~\cite{Khachatryan:2014dka}.
Nevertheless, the peak may be explained in leptoquark models with colorons~\cite{Bai:2014xba}.

In this work we argue that the simplest and most natural common explanation for the the observed excesses could be the s-channel resonant production
of right-handed gauge  bosons 
$W_R$~\cite{Keung:1983uu,Ferrari:2000sp,Rizzo:2007xs,Frank:2010cj,Schmaltz:2010xr,Nemevsek:2011hz,Grojean:2011vu,Cao:2012ng,deBlas:2012qp,Han:2012vk}, 
with right-handed gauge coupling \mbox{$g_R\sim0.6 g_L$}. The excess in $eejj$~\cite{Khachatryan:2014dka} is then produced by the subsequent decay of the $W_R$ to a heavy right-handed electron neutrino and an electron. 
The right-handed neutrino is unstable and undergoes a 3-body decay to two light jets and another electron, so the decay chain is:
\[  W_R \to N_R + e, \qquad N_R \to e + 2j . \]
Within this model it can be easily explained why the $W_R$ only decays to electrons and not to muons,
if the right-handed muon neutrino is not kinematically accessible. Indeed, as explained below, the natural 
mass scale of the symmetry breaking sector is 2-3 times the $W_R$ mass, so
that such a mass hierarchy not entirely unexpected\footnote{While this work was under preparation, similar claims for 
the $W_R$ properties were put forward in~\cite{Deppisch:2014qpa}. However, that paper does not consider the $e\nu jj$ excess
nor study other tests of the $W_R$ scenario that are the purpose of this work.}.
At the same time, the produced $W_R$-bosons are also expected to decay hadronically through the decay chain
\[ W_R \to t b/\, t q \to \ell\nu bb/ \,\ell\nu bq,\] 
where the decay rates to different quark flavours $b,s,d$ depend on the unknown values of the right-handed CKM matrix $V_R^{CKM}.$ 
Since  the search~\cite{Khachatryan:EXO12041} does not make use of \mbox{$b$-tagging}, it is sensitive to all the above final states, explaining the excess. 

If this interpretation is correct, the discovery of right-handed currents and the associated physics  at energies accessible at the LHC
would completely change our presently favoured picture of particle physics.

The asymmetric nature of elementary particle interactions has been a long-standing puzzle, dating back to
the original discovery of parity violation in kaon decays. In the Standard Model this asymmetry 
is represented by the absence of weak couplings to right-handed fermions. At the same time, the 
peculiar anomaly free chiral representations of fermions in the Standard Model point to a 
unifying structure at a higher energy scale.

These observations led Pati and Salam~\cite{Pati:1974yy} to propose the first model of partial unification based on the gauge group 
$SU(4) \times SU(2)_L \times SU(2)_R$ that is left-right symmetric. Soon after that the left-right symmetric model based on 
$SU(3)_{QCD} \times U(1)_{B-L} \times SU(2)_L \times SU(2)_R$ was formulated~\cite{Mohapatra:1974hk,Mohapatra:1974gc,Senjanovic:1975rk}.
The latter can naturally explain the smallness of active neutrino masses that are suppressed by the high $SU(2)_R\times U(1)_{B-L}$ 
breaking scale~\cite{Mohapatra:1979ia}.  
Consequently,  the baryon asymmetry of the Universe can be generated via leptogenesis~\cite{Fukugita:1986hr}.

Those models are in good accordance with the Grand Unification paradigm since they can be embedded into the
(left-right symmetric) $SO(10)$ gauge group,
\begin{align*} 
& SO(10) \\
\supset \; & SU(4) \times SU(2)_L \times SU(2)_R \\
\supset \; & SU(3)_{QCD} \times U(1)_{B-L} \times SU(2)_L \times SU(2)_R \\
\supset \; & SU(3)_{QCD} \times SU(2)_L \times U(1)_Y \\
\supset \; & SU(3)_{QCD} \times U(1)_{QED}  .
\end{align*}
Combining the Standard Model fermions plus right-handed neutrinos into a complete {\bf 16} representation of $SO(10)$ guarantees
the absence of gauge anomalies, and is also motivated by String Theory arguments. 
The $SO(10)$ symmetry breaking down to QCD and QED may occur in different chains via different intermediate scales. 
In the simplest  left-right symmetric models with one intermediate scale the gauge coupling unification implies a very high scale
of $SU(2)_R$ breaking~\cite{Deshpande:1992au}, in agreement with small neutrino masses and leptogenesis.
In more involved models,  in which the discrete left-right symmetry and the $SU(2)_R$ breaking occur at different scales~\cite{Chang:1983fu},
the $SU(2)_R$ breaking scale can be as low as the TeV scale~\cite{Deppisch:2014qpa}.

The scenario described above is appealing both phenomenologically and theoretically. 
However, no experimental evidence for the existence of right-handed currents has been obtained so far.
Constraints from precision data~\cite{Hsieh:2010zr,delAguila:2010mx} and 
from $K-\bar K$ and $B-\bar B$ systems require the $W_R$ gauge boson to be heavier than about 
2-3~TeV~\cite{Beall:1981ze,Barenboim:1996yg,Zhang:2007da,Maiezza:2010ic,Bertolini:2014sua}.\footnote{These bounds are subject to large uncertainties in low energy hadronic matrix elements and can also be relaxed by giving up assumed exact left-right symmetry $g_L=g_R,$ $V_L^{CKM}=V_R^{CKM}$~\cite{Barenboim:1996nd}.}
The Higgs sector of left-right symmetric models~\cite{Gunion:1989in,Deshpande:1990ip,Barenboim:2001vu} offers additional tests of this scenario.
The most promising of those at the LHC is the search for doubly charged Higgs boson~\cite{Huitu:1996su}. Present experimental bounds on the mass of this particle
 from the LHC~\cite{Chatrchyan:2012ya,ATLAS:2012hi} are all below the TeV scale.
Therefore the right-handed symmetry breaking scale 6-7~TeV indicated by the LHC~\cite{Khachatryan:2014dka} is safely above the existing constraints.

In the following we will propose and study additional tests of the right-handed currents and the associated physics at the LHC.
This includes searches for new resonances related to the extended gauge and Higgs sectors as well as new observables, such as asymmetries,
that are sensitive to right-handed currents. Clearly, our motivation is to point out that powerful cross-checks of this scenario  can and should be 
carried out at the LHC.
We take a bottom up approach and study the above described physics as model independently as possible just relying on group theory and on the LHC results.
Nevertheless we show that some quantitatively robust conclusions can be drawn which allow further tests of this scenario at the LHC.

\section{Gauge Coupling Strength}

From group theory arguments and from the observed strength of the strong and hypercharge interactions we can find
a simple estimate for the expected strength of the $W_R$ coupling. At the scale of \mbox{$SU(4)\to SU(3)_{QCD}\times U(1)_{B-L}$}
symmetry breaking the gauge couplings should be identical.
We can represent the $SU(4)$ generators by $4 \times 4$ matrices in 3+1 block diagonal form: 
\be
 T^{QCD}_i = \frac12 \, \begin{pmatrix} \lambda_i & 0 \\ 0 & 0 \end{pmatrix},
\quad T^{B-L} = \sqrt{\frac32} \begin{pmatrix} 1/6 & 0 \\ 0 & -1/2 \end{pmatrix} .
\ee
Here $\lambda_i$ represent the Gell-Mann matrices, and the normalization constant has been
chosen such that $ \mbox{tr} ( T_i T_j ) = \frac12 \delta_{ij}$. From this argument we find that, 
at the scale of $SU(4)$ symmetry breaking,
\begin{equation} g_{B-L} = \sqrt{\frac32} \; g_3 . \label{eq:gbml} \end{equation}
Furthermore, it is easy to see that at the scale of \mbox{$SU(2)_R\times U(1)_{B-L}\to U(1)_Y$} symmetry breaking, 
\be
 \frac1{g_Y^2} = \frac1{g_R^2} + \frac1{g_{B-L}^2}.
\ee
From this simple mix of top-down and bottom-up argument~\cite{Schmaltz:2010xr}, neglecting the logarithmic running between the two scales, we can see that if $g_{B-L} \geq 1$ as suggested by equation (\ref{eq:gbml}), the right-handed gauge coupling must be aproximately equal to the hypercharge coupling:
\be g_R \approx 0.6 \, g_L .
\ee
Curiously, as was also pointed out in~\cite{Deppisch:2014qpa}, this value fits very well to the observed signal strength associated with the $eejj$ excess~\cite{Khachatryan:2014dka}. Indeed, the CMS result excludes the $W_R$ with $g_R=g_L$ as an explanation for the excess, as the predicted signal strength in that case is larger than what is observed.

\section{Scalar sector, neutrino masses, leptogenesis}

If the gauge group just above the TeV scale is $U(1)_{B-L} \times SU(2)_L \times SU(2)_R$ as we assume here,
the minimal Higgs sector that breaks $U(1)_{B-L} \times SU(2)_R$  down to the hypercharge must contain
at least one right-handed triplet with $Y=2$ which gives mass to the $W_R$, to right-handed neutrinos $N_i$ and to itself via its VEV. 
Usually the Higgs sector is taken to be left-right symmetric~\cite{Gunion:1989in,Deshpande:1990ip,Barenboim:2001vu} containing also a
left triplet with very small VEV. In the latter scenario the Standard Model neutrinos receive mass contributions from two sources, from the usual
seesaw mechanism involving heavy neutrinos and at tree level from the left-handed triplet VEV. If the latter dominates, the doubly charged triplet component 
branching fractions to same-charge leptons must follow the measured neutrino mass matrix~\cite{Kadastik:2007yd}. 
Searches for doubly charged Higgses at the LHC14 provide good tests of this model.

According the the LHC result,  the right-handed electron neutrino $N_e$ is somewhat lighter than $N_{\mu,\tau}$. 
In this case flavoured~\cite{Davidson:2008bu} resonant~\cite{Covi:1996wh,Flanz:1996fb,Pilaftsis:1997jf} leptogenesis from the degenerate $N_{\mu,\tau}$ 
pair is possible if either a $\mu$ or $\tau$ asymmetry is generated. The latter are not washed out by $N_e$, producing the observed baryon asymmetry.
However, according the the LHC results~\cite{Khachatryan:2014dka}, only one same-charge lepton pair was observed out of 14 signal events.
If the heavy neutrino $N_e$ was of Majorana type, this ratio should have been 50:50~\cite{delAguila:2008cj}. This sets strong constraint on the nature of $N_e$ 
favouring (pseudo) Dirac heavy neutrinos~\cite{delAguila:2008hw} suggesting that the most minimal model, perhaps, is not realised in nature. 
This implies a non-minimal Higgs sector that, perhaps, can be tested at the LHC14.
This type of model building is beyond the scope of this paper.

\section{Experimental tests and existing limits}

\subsection{Resonances} 

In addition to the search described in~\cite{Khachatryan:2014dka}, other direct searches that are sensitive to $W_R$ production have been performed by the ATLAS and CMS collaborations:
\begin{enumerate}
\item
$W_R \to jj$: If the $W_R$ is produced at the LHC, it will decay to 2 jets and appear as a dijet resonance. The strongest limit on this decay was found by ATLAS using 20.3 fb$^{-1}$ at 8 TeV \cite{Aad:2014aqa}, 
excluding a ``Sequential Standard Model'' (SSM) $W'$ with a mass of 2.45 TeV at 95\% CL. However, if the $W_R$ coupling is reduced to $g_R \approx 0.6 g_L$ the mass exclusion limit drops to $M(W_R) \gsim 2$ TeV.
\item
$W_R \to tb, tj$: In this channel there are two relevant searches from ATLAS \cite{ATLAS-CONF-2013-050} and CMS \cite{Chatrchyan:2012gqa}. The 95\% CL exclusion limits from those two searches for a $W_R$ with $g_R=g_L$ are 1.84 TeV and 1.85 TeV, respectively. 
\end{enumerate}

The CMS collaboration has compared the four particle invariant mass spectrum of the $pp\to ee jj$ process to the Standard Model and to a $W_R$ model 
with different masses for the $W_R$ and the right-handed neutrino $N_R$~\cite{Khachatryan:2014dka}. In order to further test this scenario,
two easy steps could be taken in the analysis of this final state:
\begin{enumerate}
\item 
In addition to the $W_R$ mass, the mass of the right-handed neutrino can be reconstructed by measuring the invariant mass of the $\ell j j$-system.
Depending on the mass splitting between the $W_R$ and the heavy neutrino, the lepton from the original $W_R$ decay is usually expected to be more energetic. Therefore a first attempt can be to
use the lepton with less $p_T$ in the reconstruction. If the $W_R \to \ell + N_R$ hypothesis holds,
a clear peak in the $N_R$ mass distribution should be seen in signal events.
\item 
The $p_T$ distribution of the harder electron in the event should be examined. Again,
if the $W_R$ decay hypothesis is correct, a clear peak near $M_{W_R}-M_N$ should be visible.
\end{enumerate}

The symmetry breaking $SU(2)_R \times U(1)_{B-L} \to U(1)_Y$ has the same structure as the electroweak symmetry breaking in the
Standard Model, so one also expects to find an uncharged $Z_R$ boson. Parametrically, the
gauge boson masses are of the order
\be
M_{W_R}^2 \approx g_R^2 \, f^2, \qquad M_{Z_R}^2 \approx (g_R^2 + g_{B-L}^2) \, f^2,
\ee
where $f$ is the symmetry breaking scale and the exact coefficients depend on the details 
of the Higgs sector. If $g_{B-L}^2 \gsim 1$, the mixing angle between the gauge groups is 
close to maximal. The $Z_R$ boson mass is expected to be close to the symmetry breaking scale $f$, 
which is approximately 6-7 TeV if the $W_R$ lives at 2-2.5 TeV. Scalar states associated with the Higgs mechanism 
will also naturally be expected at or near this scale.
The $Z_R$ couplings are close to $B-L$ because of the large mixing angle, such that the predominant
decay channels of the $Z_R$ are to leptons. If such a $Z_R$ can be produced at the LHC14, a signal in the dilepton channels could be visible. 

The dilepton signal $Z_R \to \ell^+ \ell^-$ is also the most significant existing limit for the $Z_R$. ATLAS \cite{Aad:2014cka} and CMS \cite{Chatrchyan:2012oaa} have performed searches for this signature. The exact exclusion limits depend on the $Z_R$ charges but lie in the ballpark between 2.5-3 TeV.

\subsection{Asymmetries and Indirect Searches}

Finally, there are indirect limits on the $W_R$ model from electroweak precision (EWP) measurements, most importantly $e^+e^- \to e^+e^-$ scattering at \mbox{LEP II}.  Contributions to EWP observables are dominated by $Z_R$ exchange, while the contributions from $W_R$ and the Higgs sector are of secondary importance. The details depend on the Higgs sector, but in the limit of $g_R \approx g'$ the limit on the $SU(2)_R \times U(1)_{B-L} \to U(1)_Y$ symmetry breaking scale $f$ becomes approximately 3 TeV, implying $M(W_R)>0.9$ TeV. For a more complete discussion of EWP limits on $W_R$ models we refer the reader to \cite{Schmaltz:2010xr,Hsieh:2010zr,delAguila:2010mx}.

\begin{figure}
\begin{center}
\includegraphics[width=0.2\textwidth]{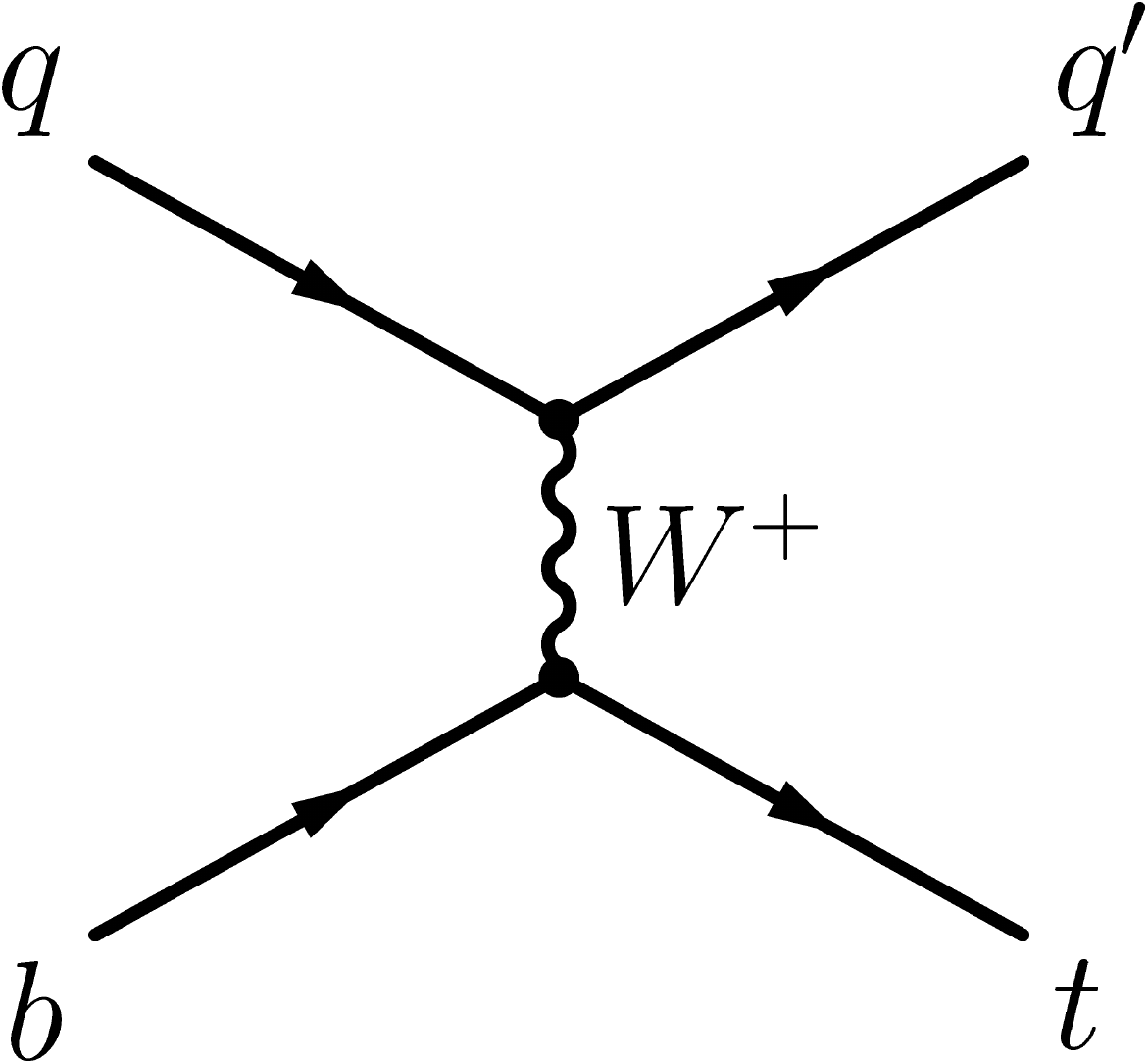}
\caption{The $t$-channel diagram contributing to single top production. In the SM, only lefthanded tops are produced via the exchange of $W_L$. In the presence of $W_R$, also right-handed top quarks can be singly produced. If the CKM element between the first and third generation in the right-handed sector is not negligible, the process can also be initiated by two light quarks, avoiding the suppression from the initial state $b$-quark PDF.}
\label{singletop}
\end{center}
\end{figure}

Because of the coupling of the $W_R$ to light quarks and to top+bottom, there will also be 
a $t$-channel contribution to single top production, competing with the left-handed $W$-exchange~\cite{Willenbrock:1986cr}, 
as shown in Figure~\ref{singletop}. Since the coupling is about 0.6 times the left-handed coupling and the $W_R$-mass is a factor of 25 higher
than the $W$-mass, we would at first expect this contribution to be insignificant. However,
if the mixing in the right-handed sector between 1st and 3rd generations is much stronger 
than in the left-handed sector, there could be a clear signal, because in that case the process can be initiated by two light quarks, and is therefore not suppressed by the PDF of the initial state b-quark.

Angular distributions in 
single top decay might then show deviations from Standard Model expectations, 
because a significant component of right-handed top quarks could be produced, contrary to the SM case~\cite{Jezabek:1994zv}. 
Searching for asymmetries~\cite{Mahlon:1996pn,Mahlon:1999gz}, \ie relations of cross sections such as $\sigma(pp\to t_L+j)/\sigma(pp\to t+j),$ 
can be an efficient way to observe the presence of new physics~\cite{Tait:2000sh,Cao:2007ea,AguilarSaavedra:2010nx,Zhang:2010dr,Aguilar-Saavedra:2014eqa}, 
as these measurements are free of systematic uncertainties such as the overall cross section normalization.

\section{Conclusions}

Recent experimental searches performed at the LHC show $\sim\!\! 3\sigma$ excesses compatible with right-handed currents mediated by a $W_R$-boson with a mass of 2-2.5 TeV. The coupling
strength of this $W_R$ appears to be about what is expected from theoretical arguments. 
Many details of this model have already been discussed, e.g. in \cite{Schmaltz:2010xr}.
Further tests are needed to confirm or falsify the $W_R$ hypothesis. 
Among those tests are simple kinematic distributions, and asymmetries which could be measured
in single top production at the LHC.

\mysection{Acknowledgment}
This work was supported by  grants MTT60, MJD435, IUT23-6, and by EU through the ERDF CoE program.

\end{document}